\documentclass[10pt]{iopart}
\usepackage{epsfig}
\usepackage{graphicx}% Include figure files
\newcommand{\be}{\begin{eqnarray}}
\newcommand{\ee}{\end{eqnarray}}

\newcommand{\ave}[1]{\left\langle #1 \right\rangle}

\begin{document}\hbadness=10000
%\topmargin=-.7cm\oddsidemargin = -0.7cm\evensidemargin = -1.7cm
%\draft
%%%%%%%%%%%%%%%%%%%%%%%%%%%%%%%%%%%%%%%%%%%
\title{Resonances and fluctuations of strange particle in 200 GeV Au-Au collisions}
\author{Giorgio Torrieri}
\address{
Department of Physics, McGill University, Montreal, QC H3A-2T8, Canada}

\date{June, 2006}

\begin{abstract}
We perform an analysis of preliminary data on strange particles yields and fluctuations  within the Statistical hadronization model.
We begin by describing the theoretical disagreements between different statistical models currently on the market.
We then show how the simultaneous analysis of yields and fluctuations can be used to differentiate between the different models, and determine if one of them can be connected to underlying physics.
We perform a study on a RHIC 200 GeV data sample that includes stable particles, resonances, and the event-by-event fluctuation of the $K/\pi$ ratio.
We show that the equilibrium statistical model can not describe the fluctuation, unless an unrealistically small volume is assumed.  Such small volume
then makes it impossible to describe the total particle multiplicity.
The non-equilibrium model,on the other hand, describes both the $K/\pi$ fluctuation and yields acceptably due to the extra boost to the $\pi$ fluctuation provided by the high pion chemical potential.
$\Lambda(1520)$ and $K^*$ abundance is described within error bars, but the $\Sigma^*$ is under-predicted to $\sim$ 1.5 standard deviations.
We suggest further measurements that have the potential to test the non-equilibrium model, as well as gauge the effect of re-interactions between hadronization and freeze-out
\end{abstract}
%%%%%%%%%%%%%%%%%%%%%%%%%%%%%%%%%%%%%%%%%%%%

\maketitle
\section{Statistical model descriptions of particle yields}
One of the main objectives of heavy ion physics is to study the collective properties of strongly interacting matter.    It's equation of state, transport coefficients,
degree of equilibration and phase structure, and the dependence of these on energy and system size.

Thus, the natural approach to study soft particle production in heavy ion collisions is through statistical mechanics techniques.   
Such an approach has a long and illustrious history \cite{Fer50,Pom51,Lan53,Hag65}.      However, the systematic quantitative comparison of data to the statistical model is a comparatively
recent field;   A consensus has developed that the statistical hadronization model can indeed fit most or all particles for AGS,SPS and RHIC energies
\cite{jansbook,bdm,cleymans,barannikova,gammaq_energy,gammaq_size,becattini}.

This consensus, is, however, somewhat superficial.  The models cited in the references above are considerably different w.r.t. each other, beyond their common origin
in statistical mechanics.   As a result, linking fit results to physics in a model-independent way has proven to be somewhat problematic.

For instance, the observation of ``the horn'', a sharp dependence of certain observables, notably the $K^+/\pi^+$ ratio, with energy \cite{horn} has generated a lot of theoretical interest, and is generally
seen as remarkably suggestive of a phase transition.
However, many completely different explanations were given for this phenomenon, all to some extent based on statistical mechanics:

The original $K^+/\pi^+$ ``horn'' dependence was predicted \cite{horn_theory} as a signature of deconfinement;  The motivating argument was that  the onset of the phase transition should
be marked by a sharp change in the strangeness over entropy ratio, tracked by $K/\pi$.
This idea has recently been realized \cite{gammaq_energy} in terms of non-equilibrium at freeze-out:  The extra strangeness and entropy in the high-temperature
phase translate into lack of chemical equilibrium \textit{after} the phase transition, resulting in super-cooling (freeze-out temperature
$\sim$ 140 MeV) and phase-space over-saturation (over-abundance of both light and strange quarks w.r.t. equilibrium expectations).
Other authors, however, have interpreted the horn without advocating deconfinement;   Models have been developed where the sharp structure arises due to 
the transition to the Grand Canonical limit for strangeness or from freeze-out conditions where mesons rather than baryons carry the bulk of the entropy and conserved quantum numbers \cite{hornthermal}.
Another proposed explanation relies on the {\em absence} of equilibration between Kaons and pions \cite{tomasik}.
While this variety of different models is a sign of both the theoretical and experimental vitality of our field, it would be very useful to rule some of these models out.

A similar ambiguity is present in the interpretation of strangeness enhancement, now firmly observed both at SPS \cite{stenhsps} and RHIC
\cite{stenhrhic} energies.     It was originally proposed on the grounds that a perturbative gluon-rich QGP is a lot more efficient as a strangeness production mechanism 
than a hadron gas \cite{jansbook}.     However, a considerable amount of recent attention has been given to models where the enhancement is explained by
a \textit{suppression} of strange particles in low-multiplicity systems due to deviation from the Grand-Canonical limit
\cite{strangeness_canonical}.
The continued ambiguity over these models is evidence that further observables need to be accessed before either of the strangeness production mechanisms can be considered ``falsified''.

Finally, a topic of considerable interest is the relationship between the statistical model and freeze-out.
The RHIC HBT puzzle \cite{hbt,hydroheinz} makes it clear that freeze-out conditions are, at the moment, not well understood.   In particular, it is unclear what, if any, is the effect of the evolution between hadronization (the formation of hadrons as effective degrees of freedom) and freeze-out (the moment when all hadrons decouple).
As shown in \cite{hydroheinz}, the discrepancy with experiment would lessen if chemical and kinetic freeze-out were to coincide.
While, as claimed in \cite{hydroheinz}, such a high freeze-out temperature would spoil agreement with particle spectra, fits based on a single freeze-out model prescription have shown 
that, provided a correct treatment of resonances is maintained, particle spectra are compatible with simultaneous freeze-out
\cite{florkowski,us_sps,us_rhic}.   This somewhat invalidates the claims  \cite{stagedfo} that ``blast-wave'' fits to spectra (where resonance feed-down is not included) are evidence that freeze-out occurs after a long
period of re-interaction, with $\Xi$ s and $\Omega$ s freezing out before $\pi,K$ and protons.

A logical step to clarify these scenarios is to directly measure the abundance of resonances.  Here, the situation becomes even more ambiguous:
As pointed out in \cite{usresonances}, resonance abundance generally depends on two quantities:  $m/T$, where $m$ is resonance mass and the chemical freeze-out temperature, as well as $\tau \Gamma$, where $\Gamma$ is the resonance width and $\tau$ is the reinteraction time.  Observing two ratios where the two particles have the same chemical composition, but different $m$ and $\Gamma$, such as $\Lambda(1520)/\Lambda$ vs $K^*/K$, or $\Sigma(1385)/\Lambda$ vs $K^*/K$, could therefore be used to extract the magnitude of the freeze-out temperature and the re-interaction time.
Studies of this type \cite{fachini,salur} are still in progress;    
As we will show, $\Lambda(1520)/\Lambda$ and $K^*/K$ seem to be compatible with sudden freeze-out, provided freeze-out happens in a super-cooled over-saturated state, as suggested in \cite{jansbook,gammaq_energy,gammaq_size}.
Other preliminary results, such as $\rho/\pi$, $\Delta/p$ \cite{barannikova}, and now $\Sigma^*/\Lambda$ \cite{salur} seem to be produced \textit{in excess} of the statistical model \cite{fachini,barannikova}, both equilibrium and not.     It is difficulty to see how a long re-interacting phase would produce such a result:  Resonances whose interaction cross-section is small w.r.t. the timescale of collective expansion would generally be depleted by the dominance
of rescattering over regeneration processes at the detectable (on-shell) mass range. More strongly interacting resonances would be re-thermalized at
a smaller, close to thermal freeze-out temperature.   Both of these scenarios would generally result in a suppression, rather than an enhancement, of directly detectable resonances.    Transport model studies done on resonances generally confirm this \cite{urqmd1,urqmd2}

The observation of an enhanced $\mu^+ \mu^-$ continuum around the $\rho$ peak \cite{na60} has been pointed to as evidence of $\rho$ broadening, which
in turn would signify a long hadronic re-interaction phase \cite{na60}.
The absence of a broadening in the {\em nominal} peak itself, prevents us from considering this as
the unique interpretation of experimental data.  Moreover, even a conclusive link of broadening with hadronic re-interactions would still give no indication to the length of the hadronic rescattering period.   Nor it would resolve the discrepancies pertaining hadronic resonances encountered in the previous paragraph;
The lack of modification in either mass or width, between p-p and Au-Au seen so far \cite{fachini,salur} is only compatible with the NA60 result\footnote{Unless one assumes that at freeze-out the SPS system to be very different from RHIC's.  But no soft data so far would lend support to such a hypothesis, and in all statistical models mentioned here conditions at SPS and RHIC are comparable, save for the larger SPS chemical potentials} provided the
$\rho$ is {\em very} quick to thermalize, so \cite{fachini} sees only the $\rho$ s formed close to thermal freeze-out.
  But, as argued in the previous paragraph, that under-estimates the abundance of the $\rho$ and other resonances, since the $\rho/\pi,K^*/K$, and even $\Lambda(1520)/\Lambda$ ratios point to a freeze-out temperature significantly above the 100 MeV, commonly assumed to be the ``thermal freeze-out'' temperature in a staged freeze-out scenario.

It is apparent that the data surveyed above is insufficient for a conclusive analysis.
One or more missing observables are required.   It is desirable that these observables have a different dependence on the non-equilibrium parameters $\gamma_{q,s}$ (see Eq. \ref{chemneq}), system volume, and temperature, so as to decorrelate these parameters and falsify the non-equilibrium picture in \cite{gammaq_energy,gammaq_size} and it's explanation for the ``horn'' and related discontinuities.    It is also desirable that these observables be strongly ensemble Dependant, to test the Canonical origin theory of strangeness enhancement.
Finally, a direct experimental sensitivity to the re-interaction period would be invaluable to resolve the conundrums of the preceding paragraphs.

As we will show, considering both yields and fluctuations within the same analysis could have all three requirements.

\section{Fluctuations in the statistical model}
Event-by-event particle fluctuations have elicited 
theoretical~\cite{jeonkochratios,fluctqgp1,fluctqgp2,prcfluct}
and experimental  interest~\cite{starfluct2,phefluct,supriya,spsfluct}, both as a consistency check
for existing models, as a test of hadron gas equilibration and as a way to search  for new physics.
The ability of the SHM to describe not just averages, but event-by-event multiplicity fluctuations
has however not been widely investigated, and its applicability is currently a matter of controversy.

In this work, we use the Grand Canonical ensemble (most physically appropriate for describing RHIC data \cite{prcfluct}), implemented in a publicly-available software package \cite{share,sharev2} to calculate fluctuations $\ave{(\Delta N)^2}$ and yields ($\ave{N}$).   At freeze-out, these are given by
\be
\label{yield_formula}
\langle N_i\rangle
  &=&  gV\int {d^3p\over (2\pi)^3}\, {1\over  \lambda_i^{-1} e^{\sqrt{p^2+m_{i}^{2}}/T}\pm 1}\\\langle(\Delta N_i)^2\rangle & = & \lambda_i{\partial \over
                 \partial \lambda_i} \langle N_i\rangle
\ee
The parameter $\lambda_i$ corresponds to the particle fugacity, related to the chemical potential by $\lambda_i=e^{\mu_i/T}$.  Provided the law of mass action holds, it should be given by the product of charge fugacities (flavor, isospin etc.).  It is then convenient to parametrize it in terms of equilibrium fugacities $\lambda_i^{\mathrm{eq}}$ and phase space occupancies $\gamma_i$.   For a hadron with $q (\overline{q})$ light quarks, $s (\overline{s})$ strange quarks and isospin $I_3$ the fugacity is then
\begin{eqnarray}
\label{chemneq}
\lambda_i = \lambda_i^{\mathrm{eq}}
\gamma_q^{q+\overline{q}} \gamma_s^{s+\overline{s}}
\phantom{A},\phantom{A}
\lambda_i^{\mathrm{eq}}=\lambda_{q}^{q-\overline{q}} \lambda_{s}^{s -
\overline{s}}
\lambda_{I_3}^{I_3} 
\end{eqnarray}
If the system is in chemical equilibrium then detailed balance requires that $\gamma_q=\gamma_s=1$.    Assuming $\gamma_q=1$ and fitting particle
ratios gives the $T \sim 160-170$ MeV freeze-out temperature typical of chemical equilibrium freeze-out models at SPS and RHIC \cite{bdm,cleymans,barannikova}.

In a system expanding and undergoing a phase transition, however, the condition of chemical equilibrium 
no longer automatically holds, so one has to allow for the possibility that $\gamma_q \ne 1,\gamma_s \ne 1$.
In particular, if the expanding system undergoes a fast phase transition from a QGP to a hadron gas, chemical non-equilibrium  \cite{jansbook} and super-cooling \cite{csorgo} can arise due to entropy conservation:
By dropping the hadronization temperature to $\sim 140$ MeV and oversaturating the hadronic phase space above equilibrium ($\gamma_q \sim 1.5,\gamma_s \sim 2$), it is possible to match the entropy of a hadron gas with that of a system of nearly massless partons \cite{jansbook}.
In such a case, the pion chemical potential is very close to, but not quite reaching the point of Bose-Einstein condensation $\lambda_i \rightarrow e^{m_i/T}$.

The onset of these conditions (and of the associated strangeness over-saturation condition $\gamma_s>1$) is the basis of the non-equilibrium statistical
model description of the ``horn'' mentioned in the preceding section \cite{gammaq_energy}.

This limit should result in enhanced fluctuations \cite{prcfluct}, since higher order terms in the Bose-Einstein expansion are always
more significant than for yields: for $\lambda_i  e^{m_i/T} = 1-\epsilon$ $\ave{N_i}$ converges but $\ave{\Delta N_i^2}$ diverges as $\sim \epsilon^{-1/2}$.
Thus, a measure of both fluctuations and yields could serve to falsify the non-equilibrium and equilibrium statistical model scenarios:
$T$ and $\gamma_q$ correlate in yields and anti-correlate in fluctuations.  Measuring both yields and fluctuations can be used to determine both temperature 
and $\gamma_q$ precisely.

The final state yield of particle $i$ is computed by adding the direct
yield and all resonance decay feed-downs.
\be
 \label{resoyield}
\langle N_i\rangle_{\rm total} & = &
\langle N_i\rangle + \sum_{{\rm all}\;j \rightarrow i}
B_{j \rightarrow i}  \langle   N_j\rangle \\
\ave{(\Delta N_{ i})^2}
& = & 
 \sum_{{\rm all}\;j \rightarrow i} B_{j\to i}(1-B_{j\to i})\ave{N_j}
+
 \sum_{{\rm all}\;j \rightarrow i} B_{j\to i}^2 \ave{(\Delta N_j)^2}\phantom{.....}
\ee
assuming no primordial correlation between $i$ and $j$.

The fluctuation of a ratio $N_1/N_2$
can be computed from the fluctuation of the denominator and the numerator \cite{jeonkochratios}
($\sigma_{X}^2=\ave{(\Delta X)^2}/\ave{X}$):
\begin{equation}
\label{fluctratio}
\sigma_{N_1/N_2}^2
= \frac{\ave{(\Delta N_1)^2}}{\ave{N_1}^2}
+ \frac{\ave{(\Delta N_2)^2}}{\ave{N_2}^2}
- 2 \frac{\ave{\Delta N_1 \Delta N_2}}{\ave{N_1}\ave{ N_2}}.
\end{equation}
This is a a useful experimental observable, since volume fluctuations, difficult to model from first principles, cancel out event-by-event.

Note that fluctuations of ratios scale as the inverse of the absolute normalization, $\sigma_{N_1/N_2} \sim (\ave{V}T^3)^{-1}$.
Hence, a fit including such a fluctuation needs to have absolute normalization as one of the fit parameters.
Fitting a particle multiplicity, $\sim (\ave{V}T^3)$ together with the ratio provides a way to measure the system volume at chemical freeze-out with great
precision, due to the very different scaling of the two types of observables. 

Note also the appearance of a negative correlation term  between $N_1$ and $N_2$ stemming from a common
resonance feed-down ($\Delta \rightarrow p \pi$ will be a source of correlation between $N_{p}$ and $N_{\pi}$).
\begin{equation}
\ave{\Delta N_1 \Delta N_2} \approx \sum_j b_{j \rightarrow 1 2} \ave{N_j}
\end{equation}
This correlation term is an invaluable phenomenological resource, since it is sensitive to resonance abundance at {\em chemical} freeze-out \cite{jeonkochratios};  Further reinteraction, provided resonances are not kicked out of the detector's acceptance region \cite{prcfluct}, preserve the correlation even when the original resonance
stops being reconstructible.   Thus, a measurement of the correlation together with the directly detectable resonance is a direct measurement of the amount
of re-interaction between the thermal and chemical freeze-out.

As can be seen, fluctuations measurements fulfill all three of the requirements enumerated in the last section.  They are sensitive to the degree
of chemical non-equilibrium and the system volume.   They can gauge the amount of hadronic re-interaction after hadronization.   And, as shown in \cite{nogc2},
they are strongly ensemble-specific.   In the next section, we will attempt to interpret current RHIC data in both yields and fluctuations in light of the controversies described in the previous section, and to suggest experimental observables capable of further clarifying the situation.

\section{Statistical model comparison with preliminary RHIC data}

%%%%%%%%%%%%%%%%%%%%%%%%%%
\begin{figure}[tb]
\epsfig{height=7.cm,clip=,figure=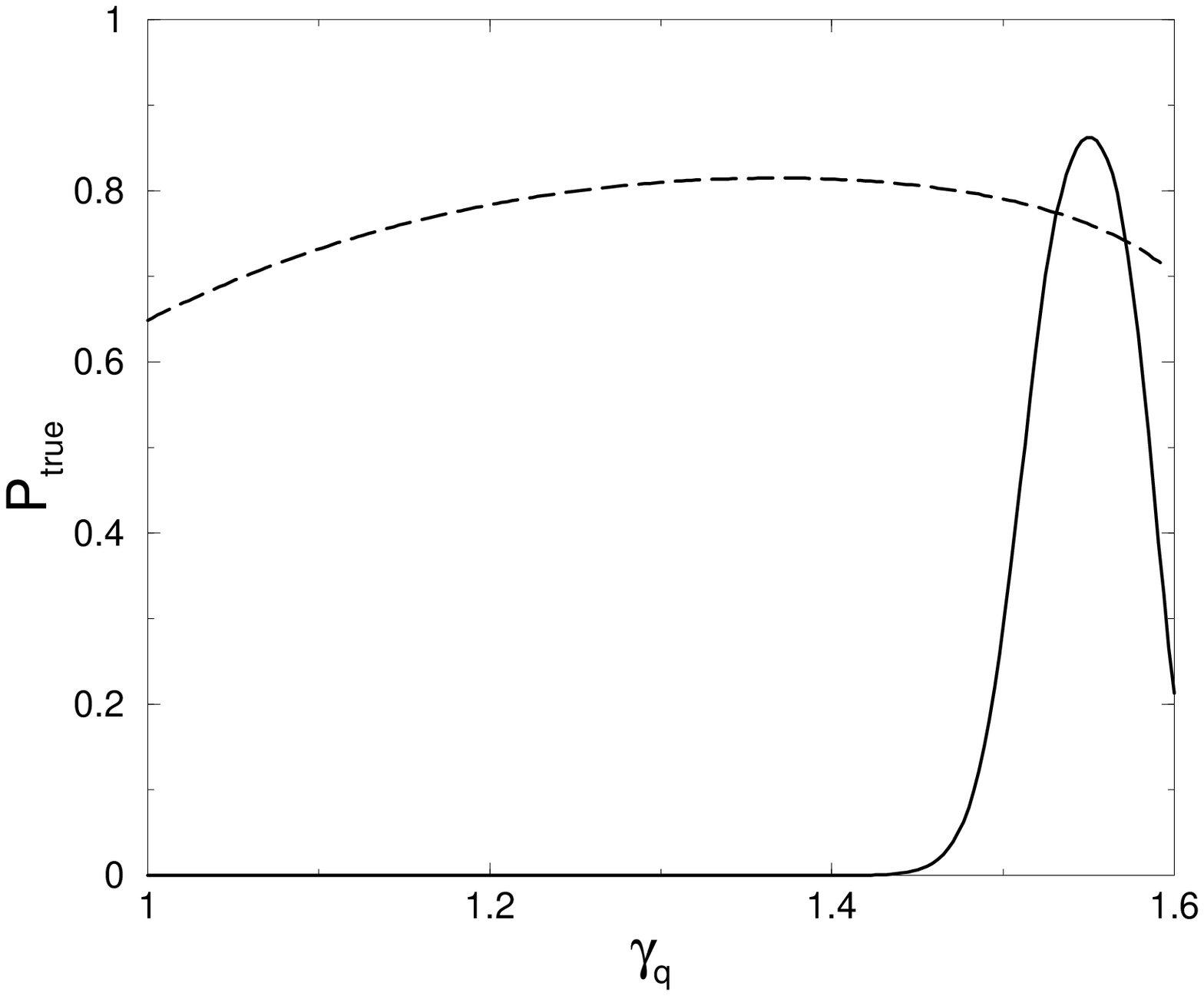}
\epsfig{height=7.cm,clip=,figure=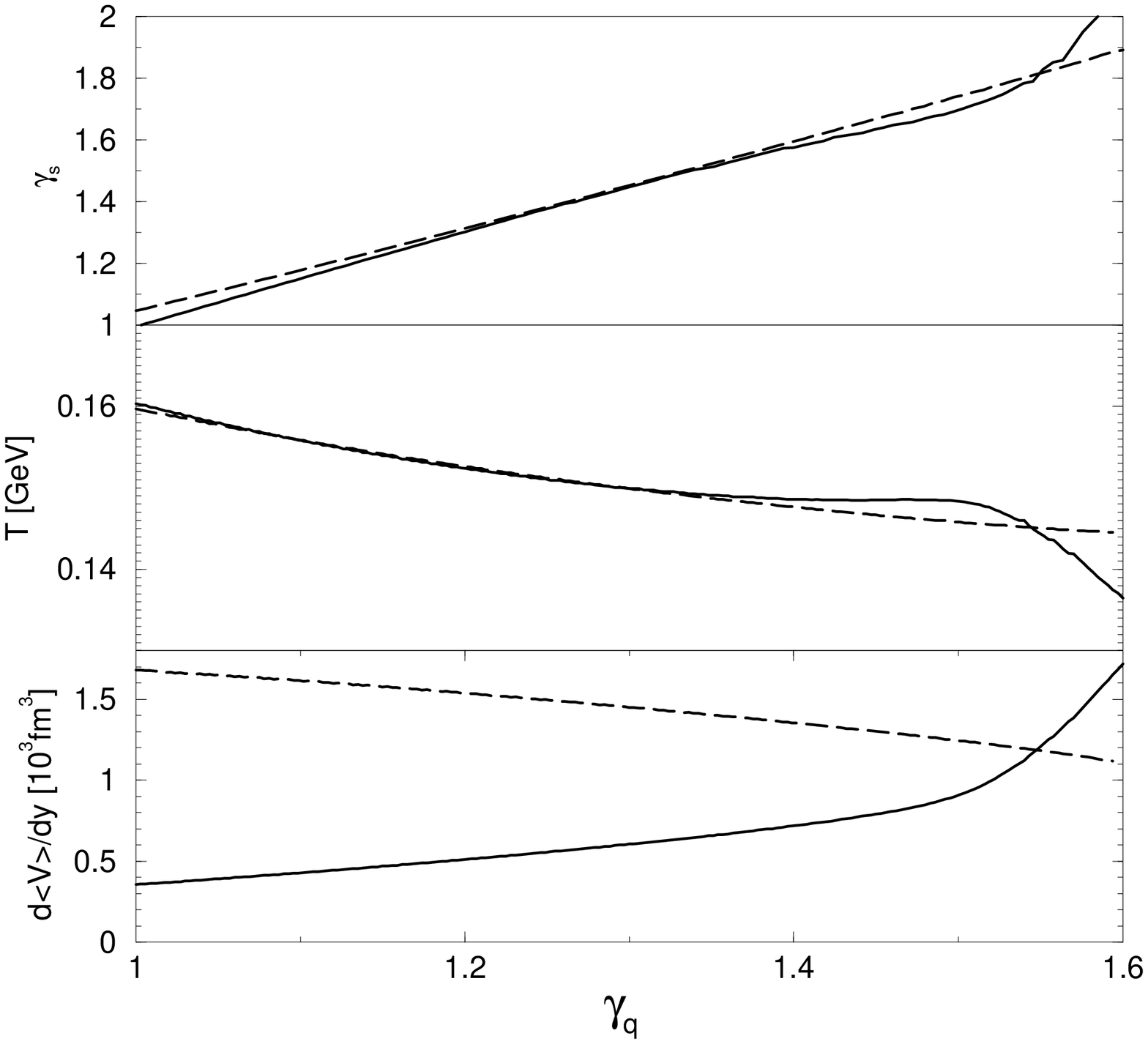}
\caption{\label{par}Left panel:  Statistical significance profile for the 200 GeV fit including (solid line) and excluding (long-dashed line) the $\sigma^{dyn}_{K/\pi}$ data-point. Right panel: Parameter sensitivities for the fits for $T$,$d\ave{V}/dy$ and $\gamma_s$} 
\end{figure}
%%%%%%%%%%%%%%%%%%%%%

The main experimental problem with fluctuation measurements is the vulnerability to effects resulting from limited detector acceptance.
This difficulty can be lessened, to some extent, by considering ``dynamical'' fluctuations, obtained by subtracting a ``static'' contribution which
should be purely Poisson in an ideal detector.
\begin{equation}
\sigma^{dyn}=\sqrt{\sigma^2-\sigma_{stat}^2}
\end{equation}
$\sigma_{stat}$, usually obtained through a Mixed event approach \cite{pruneau},
includes a baseline Poisson component, which for a ratio $N_1/N_2$ can be modeled as
\begin{equation}
\label{poissrat}
\sigma_{stat}^2 = \frac{1}{\ave{N_1}}+\frac{1}{\ave{N_2}}
\end{equation}
 as well as a contribution from detector efficiency and kinematic cuts.    Provided certain assumptions for the detector response function hold (see appendix A of \cite{pruneau}), subtracting
$\sigma_{stat}$ from $\sigma$ should yield a ``robust'' detector-independent observable.

%%%%%%%%%%%%%%%%%%%%%%%%%
\begin{figure}[tb]
%\begin{center}
\epsfig{width=7.cm,clip=,figure=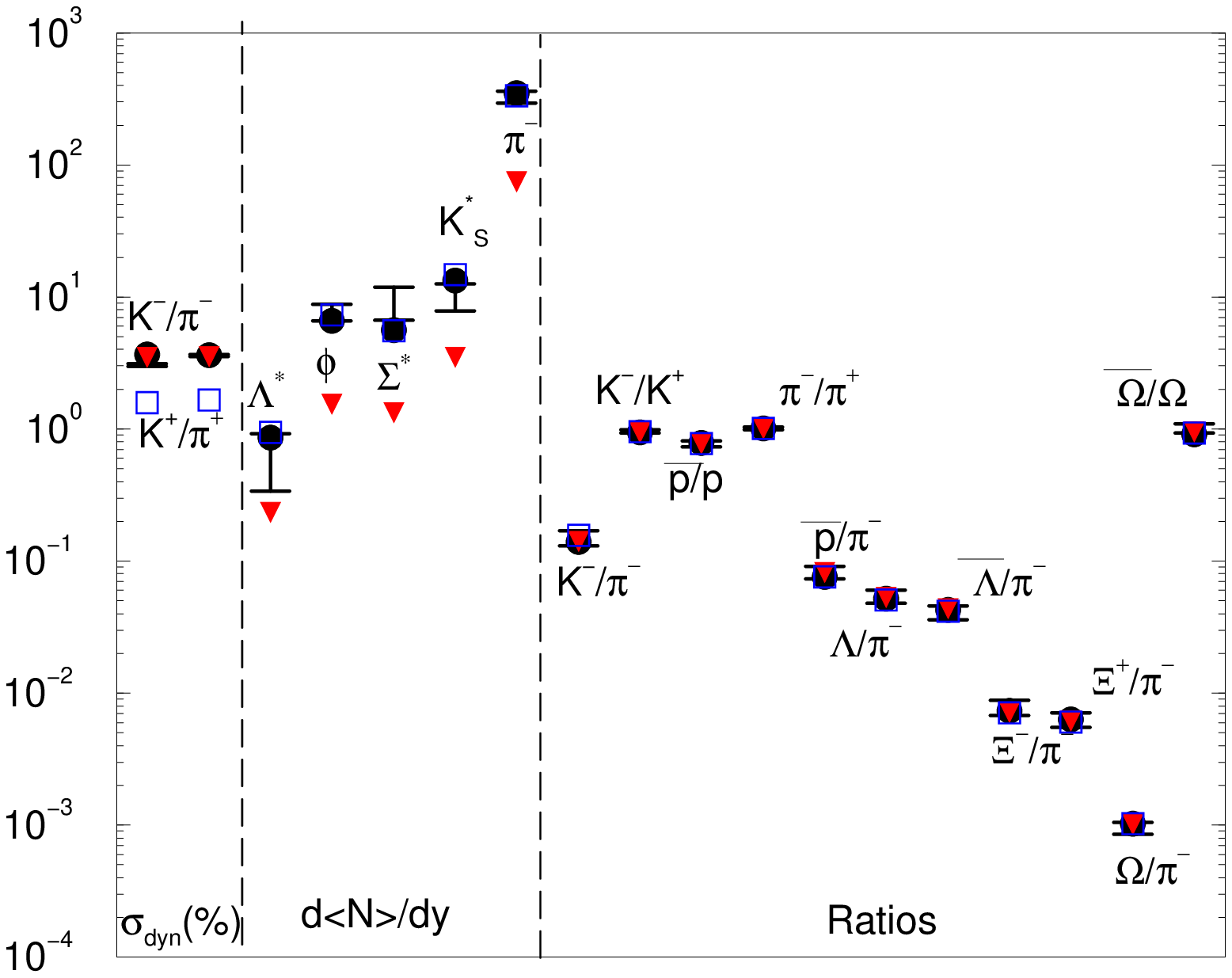}
\epsfig{width=7.cm,clip=,figure=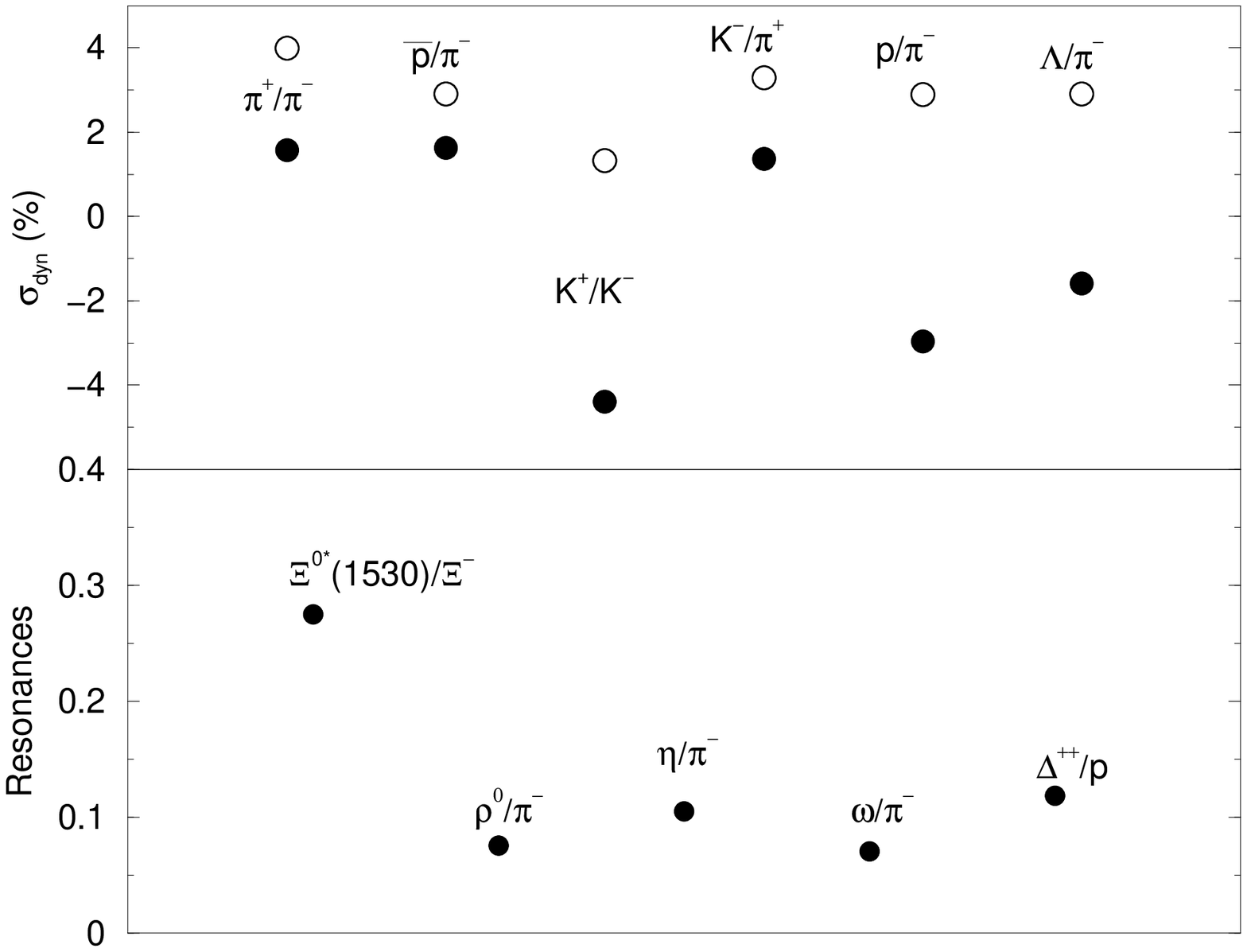}
\caption{\label{graph200}
Left panel: Fit of preliminary 200 GeV data, including the $K^{\pm}/\pi^{\pm}$ fluctuations and the $K^* (892)$ and $\Lambda^* (1520)$ resonance. Black circles represent a fit where $\gamma_q$ was fitted.  Red triangles down, $\gamma_q=1$.   Blue squares, $\gamma_q=1$ and fluctuation not fitted. Right panel:  Predictions for resonances and fluctuations calculated with best fit values.   Filled circles represent full resonance correlation.  Hollow circles assume 20 $\%$ of correlations due to resonances survive within the detector acceptance region, compatible with the calculation in
\cite{prcfluct} for $\Delta \eta=\pm 0.5$   }
%\end{center}
\end{figure}
%%%%%%%%%%%%%%%%%%%% 

%%%%%%%%%%%%%%%%%%%%%%%%%%
%\begin{figure}[tb]
%\epsfig{width=8.cm,clip=,figure=pdat_gq.eps}
%\epsfig{width=8.cm,clip=,figure=pdat_T.eps}
%\caption{\label{dat} Temperature and $\gamma_q$ data-point sensitivities for resonances and fluctuations }
%\end{figure}
%%%%%%%%%%%%%%%%%%%%%

We have performed a fit incorporating all ratios from \cite{barannikova}, with the exception of the $\Delta^{++}/p$, which the STAR collaboration has taken back in more recent analyzes \cite{starwhitepaper}.  In addition, we have included the preliminary value
for the $K/\pi$ event-by-event fluctuation measured by STAR \cite{supriya}, as well as the published yield for $\phi$ \cite{starphi200}, $\pi^-$ \cite{starpi200}, $K^*$ \cite{fachini} , $\Lambda(1520)$\footnote{This was taken at $5\%$ rather than 10$\%$ centrality.   The systematic error on normalization should not however be a dominant effect} and $\Sigma^*$ \cite{salur}.    As was pointed out in \cite{qm2005kpi}, the statistical model can not at present describe the difference between $\sigma^{dyn}_{K^-/\pi^-}$ and 
$\sigma^{dyn}_{K^+/\pi^+}$, although each of the data-points can be fitted acceptably on it's own, with a parameter variation of $10\%$ or so.
Since these data points do not at present have a systematic error estimate, we do not regard this as evidence against our model.  We choose, therefore, to fit 
the  $\sigma^{dyn}_{K^-/\pi^-}$.
Our fit parameters include the normalization (hopefully related to the system ``volume'' at chemical freeze-out), temperature, $\lambda_{q,s,I_3}$ and $\gamma_{q,s}$.
We also require, by implementing them as data-points, strangeness, charge and baryon number conservation
\[\  \ave{s-\overline{s}}=0 \pm 0.01 \phantom{AAAAA}\frac{\ave{Q}}{\ave{B}}=\left[ \frac{\ave{Q}}{\ave{B}}\right]_{Au}=0.4 \pm 0.01\]
The results of the analysis are shown in Figs  \ref{par} and \ref{graph200}.
Fig. \ref{par} (left panel) shows a statistical significance profile in $\gamma_q$.  At each point in the abscissa, a fit was performed
minimizing all the other parameters ($T,\lambda_{q,s,I3},\gamma_s,d\ave{V}/dy$).  The ordinate shows the statistical significance of such a best fit.   The right panel shows what the best fit values of each parameter are at each point in $\gamma_q$.   The left panel of Fig. \ref{graph200} shows what the best fit to data looks like in three cases: $\gamma_q$ fitted (circles), $\gamma_q=1$ (triangles), and $\gamma_q=1$ and $\sigma_{K/\pi}$ not fitted (hollow squares).

As can be seen in both Fig. \ref{par} and \ref{graph200} (left panels), incorporating fluctuations, yields and ratios rules out equilibrium and forces $\gamma_q$ to be well above
unity.    
An equilibrium fit of the particle yields and ratios gives an acceptable description of particle ratios, somewhat over-predicts the $K^*$ and $\Lambda(1520)$,
but under-predicts $\sigma_{K/\pi}^{dyn}$ by many standard deviations.  Choosing the Canonical ensemble for strangeness would only make this disagreement worse
\cite{comingpaper}.
If $\sigma_{K/\pi}^{dyn}$ is fitted together with ratios (red triangles down in Fig. \ref{graph200}), it forces the system volume $d \ave{V}/dy$ to be unrealistically small $(\sim 500 fm^3)$, thereby under-predicting particle
yields by several standard deviations.
It is only the addition of $\gamma_q$  (circles in Fig \ref{graph200}) that allows fluctuations to be driven to a high enough value while
maintaining sufficiently high volume to describe the particle multiplicities, and sufficiently high temperature to describe ratios.

The correlation between $T$ and $\gamma_q$ given by particle ratios then fixes temperature to $T=140$ MeV, as predicted in a super-cooling scenario \cite{jansbook,csorgo} and $ d \ave{V}/dy$ to $\sim 1000-1500fm^3$, in agreement with earlier estimates \cite{gammaq_size} and combatible with HBT measurements \cite{hydroheinz}.  $\gamma_s$ is
also required to be significantly above 1, also in agreement with earlier studies \cite{gammaq_energy,gammaq_size} and with a scenario where strangeness enhancement occurs because of the appearance of a phase where equilibrium strangeness content is higher than for an equilibrium hadron gas \cite{jansbook}.  Quantitative
disagreements w.r.t. \cite{gammaq_energy,gammaq_size} in $ d \ave{V}/dy$,$\gamma_s$, and chemical potentials arise from data choice, and can therefore be considered a systematic error.

It is important to underline that both yields and fluctuations contribute to such a precise determination:  Equilibrium statistical models can describe most 
yields and ratios acceptably with $\gamma_q=\gamma_s=1$, but fail to describe the event-by-event fluctuation.  Conversely, transport models
provide an acceptable description of event-by-event fluctuations \cite{flucttrans}, but fail to describe the yield of multi-strange particles \cite{urqmdstrange}.  

Unlike what is sometimes asserted, the $\Lambda(1520)$ and $K^*$ are acceptably described by the statistical model.  The strongest disagreement arises
from $\Sigma^*$ under-prediction, at the level of 1.5 standard deviations.   We await for more measurements of resonances such as $\Delta$ and $\rho$ in
central collisions before trying to interpret this under-prediction.

The acceptable description of the $\Lambda(1520)$ and $K^*$ yield using the same freeze-out temperature as the stable particles, and the {\em under-prediction}
of the $\Sigma^*$, makes a case for the proposition that the re-interaction period between hadronization and freeze-out might be not as significant as generally
thought.   However, the current data is not capable to {\em rule out} such a long-reinteraction period.

The forthcoming measurement of ratios such as $\sigma^{dyn}_{K^-/\pi^+}$ and $\sigma^{dyn}_{K^+/\pi^-}$, however, will make such a falsification possible, since, unlike in the $\sigma^{dyn}_{K^-/\pi^-}$ case, a sizable resonance correlation term exists.
For example, comparing 
\begin{eqnarray}
\fl \phantom{A}\frac{1}{2} \left[  (\sigma^{dyn}_{K^+/\pi^+})^2 - (\sigma^{dyn}_{K^+/\pi^-})^2  \right] + \frac{1}{2} \left[  (\sigma^{dyn}_{K^-/\pi^-})^2 - (\sigma^{dyn}_{K^-/\pi^+})^2  \right] = \frac{\ave{\Delta K^+ \Delta \pi^-}}{\ave{K^+} \ave{\pi^-}} + \frac{\ave{\Delta K^- \Delta \pi^+}}{\ave{K^-} \ave{\pi^+}} \nonumber \\ \approx \frac{2}{3}  \alpha_{K^{*0} \rightarrow K \pi} \left[ \frac{\ave{K^{*0} (892)}}{\ave{K^+}\ave{\pi^-}} +  \frac{\ave{\overline{K^{*0}} (892)}}{\ave{K^-}\ave{\pi^+}}  \right]_{chemical\phantom{A}f.o.}
\end{eqnarray}
with the directly measured $\frac{K^{* 0}}{K^-}$ gives the amount of re-interaction $K$s and $\pi$s went through between hadronization and freeze-out.
The biggest systematic uncertainty of this measurement is likely to come from the need to estimate $\alpha_{K^* \rightarrow K \pi}$, the probability that the decay products of $K^{*0}$ will be in the detector acceptance region provided the original resonance was.   In \cite{prcfluct} we discuss this number, and give some estimates
of it's effect.   
A more precise estimate can be obtained via the Monte-Carlo
based algorithms used to estimate the experimental acceptance as a function of
phase space for the directly detected $K^*$.
We believe that, while experimental acceptance has to be taken into account, it's effect is not strong enough to destroy the possibility of an
estimate of $\left[ \frac{K^{* 0}}{K^-} \right]_{chemical\phantom{A}f.o.}$

Fig. \ref{graph200} (right panel) develops further the reasoning above, and uses the best fit parameters obtained using the current fit to calculate both
resonances and fluctuations likely to be measured soon.  The fluctuation calculation is given with both all $\alpha_{A B \rightarrow C}=1$ for all $A,B,C$ (filled circles) and $\alpha_{A B \rightarrow C}=0.2$ for all $A,B,C$  (hollow circles) to give an estimate of the systematics arising from resonance decays within a small experimental acceptance.

Perhaps the measurement most likely to constrain the $\gamma_q >1$ scenario is $\sigma^{dyn}_{K^+/K^-}$, since this data-point is completely independent of pionic observables.
As explained in the preceding section, $\gamma_q>1$ is better able to describe
fluctuations because the high pion chemical potential gives pion fluctuations
an additional boost.  Thus, fluctuations of quantities not involving pions
should be considerably suppressed since, even at $\gamma_s \sim 2$, B-E contributions for Kaons are too small to lift Kaon fluctuations too much above the Poisson limit.   This explains why the lowest value for  $\sigma^{dyn}$ in the given sample is $\sigma^{dyn}_{K^+/K^-}$, even through resonance correlations for this ratio are not nearly as numerous as for the case of, for example $\sigma^{dyn}_{p/\pi}$.  It will shortly be clear weather the smaller  $\sigma^{dyn}_{K^+/K^-}$ is experimentally confirmed.

In conclusion, we have used preliminary experimental data to show that, at 200 GeV, the equilibrium model is unable to describe both yields and fluctuations within the same statistical parameters.
The non-equilibrium model, in contrast, succeeds in describing almost all of the yields and fluctuations measured so far at RHIC, with the parameters expected from a scenario where
non-equilibrium arises through a phase transition from a high entropy state, with super-cooling and oversaturation of phase space.   We await more published data to determine weather the non-equilibrium
model is really capable of accounting for both yields and fluctuations in all light and strange hadrons produced in heavy ion collisions.

Work supported in part by grants from
the Natural Sciences and Engineering research
council of Canada, the Fonds Nature et Technologies of Quebec, and the Tomlinson foundation.  We would like to thank J. Rafelski, C. Gale and S.Jeon for helpful discussions and continued support.

\end{document}